\documentclass{aa}
\usepackage[utf8]{inputenc}
\usepackage{graphicx}
\usepackage{txfonts}
\usepackage{lipsum}
\usepackage{subcaption}
\usepackage{lscape}
\usepackage{placeins}
\usepackage{hyperref}
\usepackage{tikz}
\usepackage{listings}
\usepackage{amsmath,amssymb}
\usepackage{booktabs}
\usepackage{comment}
\usepackage{threeparttable}
\usepackage{natbib}
\usepackage{hyperref}
\usepackage{sidecap}

\usepackage{scalerel}
\usepackage{tikz}
\usetikzlibrary{svg.path}
\definecolor{orcidlogocol}{HTML}{A6CE39}
\tikzset{
  orcidlogo/.pic={
    \fill[orcidlogocol] svg{M256,128c0,70.7-57.3,128-128,128C57.3,256,0,198.7,0,128C0,57.3,57.3,0,128,0C198.7,0,256,57.3,256,128z};
    \fill[white] svg{M86.3,186.2H70.9V79.1h15.4v48.4V186.2z}
                 svg{M108.9,79.1h41.6c39.6,0,57,28.3,57,53.6c0,27.5-21.5,53.6-56.8,53.6h-41.8V79.1z M124.3,172.4h24.5c34.9,0,42.9-26.5,42.9-39.7c0-21.5-13.7-39.7-43.7-39.7h-23.7V172.4z}
                 svg{M88.7,56.8c0,5.5-4.5,10.1-10.1,10.1c-5.6,0-10.1-4.6-10.1-10.1c0-5.6,4.5-10.1,10.1-10.1C84.2,46.7,88.7,51.3,88.7,56.8z};
  }
}
\newcommand\orcidicon[1]{\href{https://orcid.org/#1}{\mbox{\scalerel*{
\begin{tikzpicture}[yscale=-1,transform shape]
\pic{orcidlogo};
\end{tikzpicture}
}{|}}}}

\hypersetup{
    colorlinks=true,
    linkcolor=blue,
    citecolor=blue,
    filecolor=blue,
    urlcolor=blue
}

\begin{document}

\title{Optical Variability and Narrow-Line Region Kinematics in Type 2 AGNs}

   \titlerunning{Optical Variability and Narrow-Line Region Kinematics in Type 2 AGNs}
   \authorrunning{M.~Laki\'{c}evi\'{c}}

   \author{Ma\v{s}a Laki\'{c}evi\'{c}\inst{1}\thanks{\email{mlakicevic@aob.rs}}}
   \institute{Astronomical Observatory Belgrade, Volgina~7, 11060 Belgrade, Serbia}

   \date{Received December 2025}

  \abstract
   {Optical variability in Type~2 active galactic nuclei (AGNs) is rarely explored because the direct accretion-disk continuum is obscured by circumnuclear dust. Nevertheless, detectable optical variations are present in multi-epoch surveys such as SDSS Stripe~82, indicating that some component of the nuclear emission is observed indirectly, for example through scattering or partial transmission.}
   {This study explores whether this variability is statistically connected to spectroscopic parameters of the narrow-line region (NLR), using the ALPAKA catalogue of spectral measurements.}
   {A subsample of 412 Type~2 AGNs was assembled by crossmatching SDSS~Stripe~82 multi-epoch variability measurements in the $u,g,r,i,z$ bands with the ALPAKA spectroscopic catalogue. Correlations were then computed between the root-mean-square (RMS) variability amplitudes and the corresponding emission-line luminosities, kinematic widths and equivalent widths (EWs).}
   {Significant anti-correlations are found between the RMS amplitudes and [O\,III]\,4949, [O\,III]\,5007, [N\,II]\,6548 and [N\,II]\,6584 line luminosities. Velocity dispersions ($\sigma$) and EWs of forbidden-lines [O\,III]\,5007 and [N\,II]\,6584 also show moderate anti-correlations with RMS. }
   {The results demonstrate that even in obscured AGNs, optical variability carries information about the hidden nucleus. The anti-correlation between RMS and line luminosity suggests a connection between accretion stability and ionising output. Anti-correlations between RMS and the [O\,III] and [N\,II] velocity dispersions indicate a secondary correlation between optical RMS variability and the integrated kinematic state of the NLR. In addition, the anti-correlation between RMS and EW shows that the EW variations are primarily driven by changes in the continuum level, while the narrow-line flux itself remains effectively constant on the relevant timescales.

   }


   \maketitle
   \nolinenumbers

\section{Introduction}\label{sec:intro}

Optical variability is a key signature of active galactic nuclei (AGNs) and has long been used to probe accretion-disk physics \citep[e.g.][]{Ulrich1997,MacLeod2010}. In AGNs, optical variability is understood to originate from instabilities in the accretion disk and its immediate environment, including stochastic fluctuations in the mass-accretion rate and thermal reprocessing of variable high-energy emission \citep[e.g.][]{Ulrich1997,Kelly2009,Mushotzky2011}. In Type~1 AGNs, variability is observed directly, while in Type~2 AGNs it typically becomes visible through scattered or reprocessed light, or through changes in line-of-sight obscuration \citep{Goodrich1995,Tran2001}. In Type~1 AGNs, variability studies are well established through reverberation mapping \citep[e.g.][]{Peterson1993}, long-term variability statistics \citep[e.g. structure-function analyses;][]{Ulrich1997}, and stochastic models that link variability amplitude to luminosity, black-hole mass, and Eddington ratio \citep[e.g.][]{Kelly2009,Mushotzky2011}. \citet{SanchezSaez2018} found that optical variability amplitudes in AGNs anti-correlate with the Eddington ratio, indicating that more efficiently accreting black holes exhibit smaller fractional variability.

While Type~1 AGNs exhibit optical variability amplitudes an order of magnitude larger than those of Type~2 objects, approximately 11\% of Type~2 AGNs nevertheless show detectable optical variability \citep{Lawrence2018,LopezNavas2023}. Variability in obscured (Type~2) AGNs is far less explored, because the broad-line region (BLR) and accretion disk are hidden behind dusty circumnuclear material. A fraction of the nuclear light may reach the observer through scattering, leakage, or time-dependent obscuration. Variable obscuration, scattering, and related effects have been proposed as mechanisms capable of producing optical changes in obscured AGNs \citep[e.g.][]{Tran2001,LaMassa2015}.

The connection of optical variability to the narrow-line region (NLR) is largely unexplored. If the variability amplitude is indirectly linked to the ionising output of the hidden nucleus, then correlations between variability and NLR diagnostics -- such as narrow-line luminosities, equivalent widths (EWs), or velocity widths -- may still emerge at the population level. However, because the NLR is expected to trace long-term averages of the ionising emission \citep[e.g.][]{Peterson1993}, it is unclear whether long-timescale optical variability should correlate directly with NLR properties. Testing this possibility requires combining uniform multi-epoch photometry with reliable spectroscopic measurements -- an approach that has not yet been systematically applied to Type~2 AGNs.

\section{Data and methods}\label{sec:data}

Spectral parameters were taken from the ALPAKA\footnote{Downloaded from \url{https://sites.google.com/site/sdssalpaka}.} survey \citep{Mullaney2013}, which provides uniformly measured emission-line properties for 24\,264 SDSS~DR7 spectra. In ALPAKA, the emission-line fluxes are obtained from multi-component Gaussian fits to the flux-calibrated SDSS spectra and are converted to luminosities using the SDSS redshift and the standard luminosity-distance relation; for several transitions (e.g. \texttt{OIII\_5007\_LUM\_DERRED}), the luminosities are additionally corrected for internal extinction via the Balmer decrement. The corresponding full widths at half maximum (FWHM) values are derived for each Gaussian component from the same multi-component fits.

In addition to the ALPAKA-derived quantities, SDSS-based EWs and velocity dispersions ($\sigma$) included in the catalogue are also used. These values originate from the SDSS~DR7 spectroscopic pipeline, which measures single-Gaussian line widths and rest-frame EWs from the same flux-calibrated spectra. Given the spectral resolution ($R\simeq2000$) and the typical S/N in the forbidden-line region, the SDSS EW and $\sigma$ measurements are widely regarded as reliable kinematic diagnostics for SDSS-based AGN studies.

Root-mean-square (RMS; sigma-clipped standard deviation) variability amplitudes were taken from the SDSS~Stripe~82 multi-epoch photometry (column \texttt{RMS\_PSFMAG\_CLIP}), derived from the Light-Motion Curve Catalogue (LMCC) described by \citet{Bramich2008}\footnote{\url{https://das.sdss.org/value_added/stripe_82_variability/SDSS_82_public/}}. The LMCC provides homogenised $ugriz$ photometry and variability statistics for nearly four million objects, where photometric and astrometric corrections remove epoch-to-epoch systematic offsets, producing a uniform and reliable multi-epoch dataset \citep{Ivezic2007}.

From eight Stripe~82 segments (00--01 through 23--24), all sources classified as Type~2 (\texttt{AGN\_TYPE}=2) in ALPAKA were selected. After crossmatching the datasets, the final sample consists of 412 Type~2 AGNs.

For each source, the following quantities are used:
\begin{itemize}
    \item RMS variability in $u,g,r,i,z$ bands from the \texttt{RMS\_PSFMAG\_CLIP} column, which provides the sigma-clipped standard deviation of the multi-epoch point-spread-function (PSF) magnitudes. In this analysis, a noise-corrected variability amplitude is used by subtracting in quadrature the contribution of photometric measurement uncertainties, estimated from the \texttt{MEAN\_PSFMAG\_ERR\_CLIP} column, i.e.
\begin{equation}
\mathrm{RMS}_{\mathrm{cor}} = \sqrt{\max\!\left(\mathrm{RMS}_{\mathrm{PSF}}^{2} - \langle\sigma_{\mathrm{err}}\rangle^{2},\,0\right)},
\label{eq:eq1}
\end{equation}
in order to mitigate the bias that would otherwise inflate the observed RMS at low signal-to-noise (faint) flux levels and potentially induce artificial correlations with luminosity. In the remainder of the paper, this noise-corrected variability metric is referred to as $\mathrm{RMS}_{\mathrm{cor}}$. When computing correlations, objects with $\mathrm{RMS}_{\mathrm{cor}} = 0$ are excluded in the corresponding band, as they are consistent with non-detections. The quantity $\mathrm{RMS}_{\mathrm{PSF}}^{2} - \langle\sigma_{\mathrm{err}}\rangle^{2}$ is strictly positive for all objects in the $g$, $r$, $i$, and $z$ bands, implying that no sources are removed by the truncation at $\mathrm{RMS}_{\mathrm{cor}}=0$ in the bands where the main results are obtained. In the $u$ band, 8 out of 215 objects ($\sim 3.7\%$) with valid, non-zero RMS and photometric-error measurements have $\mathrm{RMS}_{\mathrm{cor}}=0$.

The relative impact of the noise correction was also examined through the ratio $\mathrm{RMS}_{\mathrm{cor}}/\mathrm{RMS}$. The median value of this ratio is very close to unity in all bands, particularly in the $g$, $r$, $i$, and $z$ bands (median $\gtrsim 0.999$), indicating that the correction has a negligible effect on the measured variability amplitudes. In the $u$ band, the median remains high (0.993), but the distribution is broader, with the 10th percentile dropping to $\sim 0.88$, indicating that a non-negligible fraction of objects is affected by photometric uncertainties. A complementary view of the impact of photometric uncertainties is provided in Appendix~\ref{appendix}, where the cumulative distributions of the $\mathrm{ERR}/\mathrm{RMS}$ ratio are shown for all bands.

    \item From the ALPAKA survey, the narrow-core components of the forbidden lines are used as the primary tracers of the NLR emission in Type~2 AGNs. Although ALPAKA also provides broad (wing) components for [O\,III] and [N\,II], their FWHMs show no significant correlations with RMS variability, and the narrow-component FWHMs likewise do not correlate with RMS. Broad H$\alpha$ and H$\beta$ components are excluded entirely, while their narrow components are not used because blending and underlying stellar absorption make them unreliable in Type~2 spectra. Accordingly, the analysis focuses on the luminosities of the forbidden narrow-core lines--[O\,III]~$\lambda\lambda5007,4959$ and [N\,II]~$\lambda\lambda6548,6584$--together with the EWs and velocity dispersions of [O\,III]~$\lambda5007$ and [N\,II]~$\lambda6584$.

\end{itemize}
All parameters were converted to logarithmic units, and outliers were removed using a $3\sigma$ clipping procedure before computing the Spearman correlation coefficients ($\rho$) and $p$ values for each combination of $\mathrm{RMS}_{\mathrm{cor},u,g,r,i,z}$ and the available spectroscopic parameters.

\section{Results}\label{sec:results}

\subsection{Correlations for Type 2 AGNs}
\label{sec:variability}

\begin{figure*}[h!]
   \centering
   \includegraphics[width=180mm]{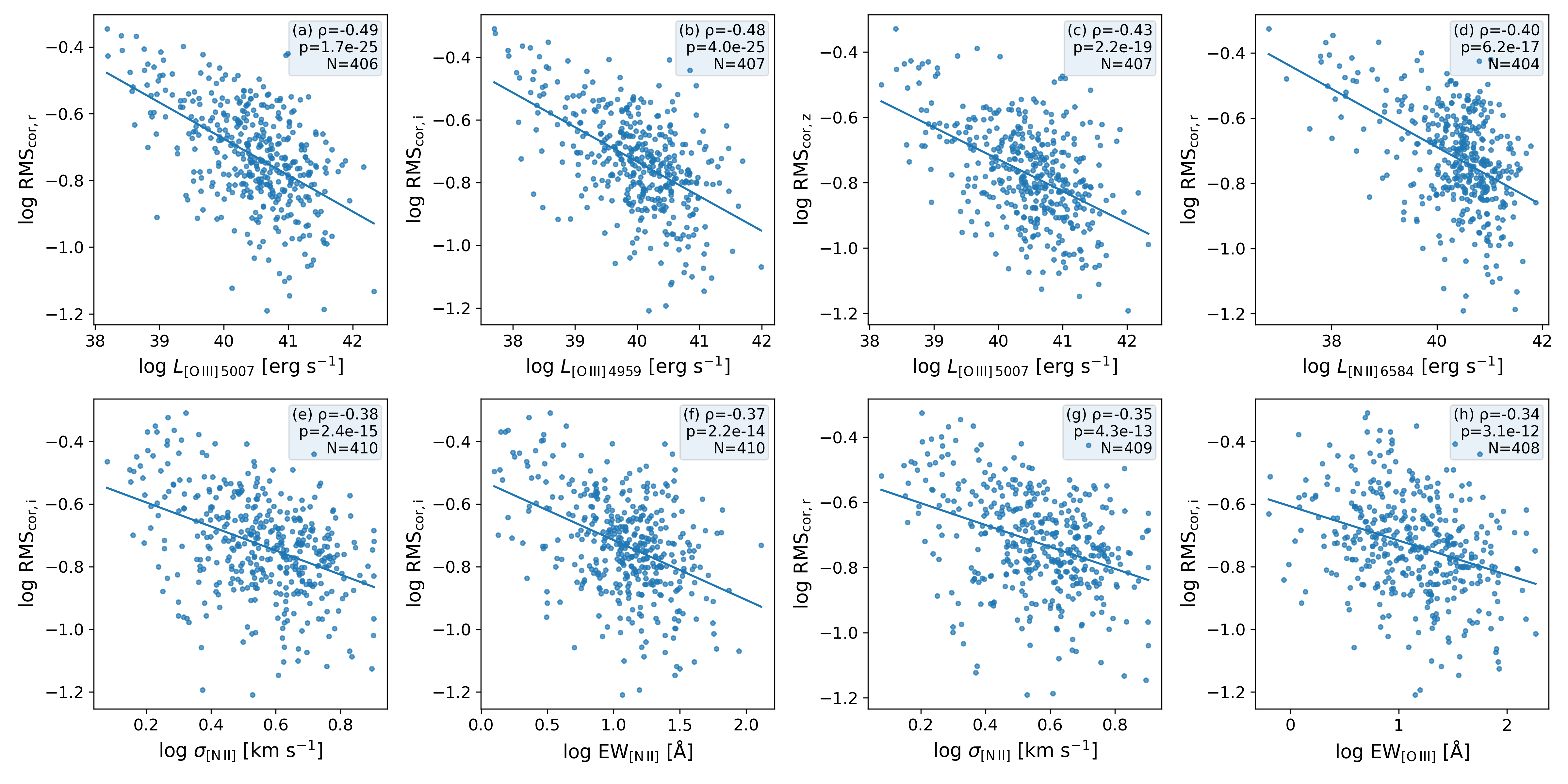}
      \caption{Representative correlations between optical RMS variability and narrow-line properties for Type~2 AGNs. Top row: anti-correlations between RMS amplitude (in the $r$, $i$, and $z$ bands) and the luminosities of [O\,III] $\lambda5007$, [O\,III] $\lambda4959$, and [N\,II] $\lambda6584$. Bottom row: corresponding correlations with the velocity dispersions and EWs of [N\,II] and [O\,III]. Each panel reports the Spearman rank coefficient $\rho$ and the associated $p$-value.}
   \label{fig:panels}
   \end{figure*}

The statistically significant correlations between $\mathrm{RMS}_{\mathrm{cor},u,g,r,i,z}$ and the [N\,II] and [O\,III] spectroscopic parameters (defined as $|\rho|>0.3$ and $p<0.05$) are listed in Table~\ref{tab:rms_trusted_corr}, with representative correlation panels shown in Fig.~\ref{fig:panels}. In all significant cases, an anti-correlation is found: sources with higher forbidden-line luminosities, larger $\sigma$, or stronger EWs tend to display lower optical RMS variability amplitudes. Because the components of the [O\,III] and [N\,II] doublets are linked by fixed atomic ratios, they are not treated as independent observables; correlations involving both components are shown only to demonstrate the robustness of the results. The strongest trends appear in the $r$, $i$, and $z$ bands. The median $\mathrm{RMS}_{\mathrm{cor}}$ does not increase toward redder bands ($u \simeq 0.19$, $g \simeq 0.17$, $r \simeq 0.16$, $i \simeq 0.15$, $z \simeq 0.14$), indicating that these stronger trends do not reflect a larger intrinsic variability amplitude at longer wavelengths. This behaviour mirrors the well-established luminosity-variability anti-correlation observed in Type~1 quasars (see Sect.~\ref{sec:discussion}).

Taken together, these three families of correlations indicate that RMS variability in Type~2 AGNs is not random with respect to NLR properties, but instead exhibits a coherent set of trends linking continuum variability, long-term ionising output, and integrated NLR kinematics.

In contrast, none of the [N\,II] or [O\,III] FWHM measurements--whether from the narrow (core) or the broad (wing) components--show statistically significant correlations with RMS variability.

\begin{table*}
\centering
\caption{Significant Spearman correlations between $\mathrm{RMS}_{\mathrm{cor}}$ (all bands) and spectral parameters for the Type~2 AGN sample.}
\label{tab:rms_trusted_corr}
\begin{tabular}{l l c c c c}
\hline\hline
RMS band & Parameter & Description & $\rho$ & $p$ & $N$ \\
\hline
RMS\_cor\_r & OIII\_5007\_LUM & Luminosity of [O\,III] $\lambda5007$ & -0.49 & 1.7e-25 & 406 \\
RMS\_cor\_i & OIII\_4959\_LUM & Luminosity of [O\,III] $\lambda4959$ &-0.48 & 4.0e-25 & 407 \\
RMS\_cor\_i & OIII\_5007\_LUM & Luminosity of [O\,III] $\lambda5007$ &  -0.48 & 8.6e-25 & 407 \\
RMS\_cor\_r & OIII\_4959\_LUM & Luminosity of [O\,III] $\lambda4959$ & -0.48 & 1.1e-24 & 406 \\
RMS\_cor\_z & OIII\_4959\_LUM & Luminosity of [O\,III] $\lambda4959$ & -0.43 & 2.0e-19 & 407 \\
RMS\_cor\_z & OIII\_5007\_LUM & Luminosity of [O\,III] $\lambda5007$ & -0.43 & 2.2e-19 & 407 \\
RMS\_cor\_r & NII\_6584\_LUM  & Luminosity of [N\,II] $\lambda6584$  & -0.40 & 6.2e-17 & 404 \\
RMS\_cor\_r & NII\_6548\_LUM  & Luminosity of [N\,II] $\lambda6548$  &  -0.40 & 6.2e-17 & 404 \\
RMS\_cor\_g & OIII\_5007\_LUM & Luminosity of [O\,III] $\lambda5007$ & -0.39 & 3.5e-16 & 405 \\
RMS\_cor\_i & NII\_6584\_LUM  & Luminosity of [N\,II] $\lambda6584$  &  -0.38 & 1.1e-15 & 405 \\
RMS\_cor\_i & NII\_6548\_LUM  & Luminosity of [N\,II] $\lambda6548$  &-0.38 & 1.1e-15 & 405 \\
RMS\_cor\_i & SDSS\_NII\_SIGMA& Velocity dispersion of the [N\,II] $\lambda6584$ & -0.38 & 2.4e-15 & 410 \\
RMS\_cor\_g & OIII\_4959\_LUM & Luminosity of the [O\,III] $\lambda4959$ & -0.38 & 5.6e-15 & 405 \\
RMS\_cor\_i & SDSS\_NII\_EW   & Equivalent width of [N\,II] $\lambda6584$ & -0.37 & 2.2e-14 & 410 \\
RMS\_cor\_r & SDSS\_NII\_EW   & Equivalent width of [N\,II] $\lambda6584$ & -0.35 & 1.6e-13 & 409 \\
RMS\_cor\_r & SDSS\_NII\_SIGMA & Velocity dispersion of the [N\,II] $\lambda6584$ & -0.35 & 4.3e-13 & 409 \\
RMS\_cor\_z & NII\_6548\_LUM & Luminosity of [N\,II] $\lambda6548$  &-0.35 & 5.8e-13 & 405 \\
RMS\_cor\_z & NII\_6584\_LUM & Luminosity of [N\,II] $\lambda6584$  &-0.35 & 5.8e-13 & 405 \\
RMS\_cor\_i & SDSS\_OIII\_EW & Equivalent width of [O\,III] $\lambda5007$ &-0.34 & 3.1e-12 & 408 \\
RMS\_cor\_r & SDSS\_OIII\_EW & Equivalent width of [O\,III] $\lambda5007$ &-0.33 & 1.2e-11 & 407 \\
RMS\_cor\_r & OIII\_5007\_LUM\_DERRED &Dereddened luminosity of [O\,III] $\lambda5007$ & -0.33 & 2.1e-11 & 402 \\
RMS\_cor\_g & NII\_6548\_LUM & Luminosity of [N\,II] $\lambda6548$ & -0.32 & 3.4e-11 & 403 \\
RMS\_cor\_g & NII\_6584\_LUM & Luminosity of [N\,II] $\lambda6584$  &-0.32 & 3.4e-11 & 403 \\
RMS\_cor\_z & SDSS\_NII\_SIGMA & Velocity dispersion of the [N\,II] $\lambda6584$ & -0.32 & 2.6e-11 & 410 \\
RMS\_cor\_i & OIII\_5007\_LUM\_DERRED &Dereddened luminosity of [O\,III] $\lambda5007$ &  -0.31 & 1.6e-10 & 402 \\
RMS\_cor\_i & SDSS\_OIII\_SIGMA &Velocity dispersion of the [O\,III] $\lambda5007$ & -0.30 & 5.2e-10 & 409 \\
RMS\_cor\_z & OIII\_5007\_LUM\_DERRED &Dereddened luminosity of [O\,III] $\lambda5007$ &  -0.30 & 8.2e-10 & 404 \\
RMS\_cor\_g & SDSS\_NII\_EW   & Equivalent width of [N\,II] $\lambda6584$ & -0.30 & 8.2e-10 & 408 \\
RMS\_cor\_u & OIII\_4959\_LUM & Luminosity of [O\,III] $\lambda4959$ & -0.30 & 1.9e-05 & 202 \\
RMS\_cor\_u & OIII\_5007\_LUM & Luminosity of [O\,III] $\lambda5007$ &-0.30 & 2.0e-05 & 202 \\
\hline
\end{tabular}

\begin{flushleft}
\textbf{Notes.} Only correlations with $|\rho|\ge 0.3$ and $p<0.05$ are listed, sorted by decreasing $|\rho|$. The column $N$ gives the number of objects used in each correlation after applying $3\sigma$ clipping.
\end{flushleft}

\end{table*}

\subsection{Control sample, non-AGN Galaxies}
\label{subsec:control-nonagn}

To assess whether the detected multi-epoch RMS variability trends in the Type~2 sample could be driven by host-galaxy systematics or observational effects, a control sample of inactive or star-forming galaxies was constructed from the same ALPAKA--Stripe~82 merged catalogue. All objects flagged as $\texttt{AGN\_TYPE}=-1$ (non-AGN) were selected, and a redshift ($z>0$) was checked to exclude stellar contaminants, yielding a relatively small control sample of 44 galaxies.

\begin{figure*}[h!]
\centering
\includegraphics[width=180mm]{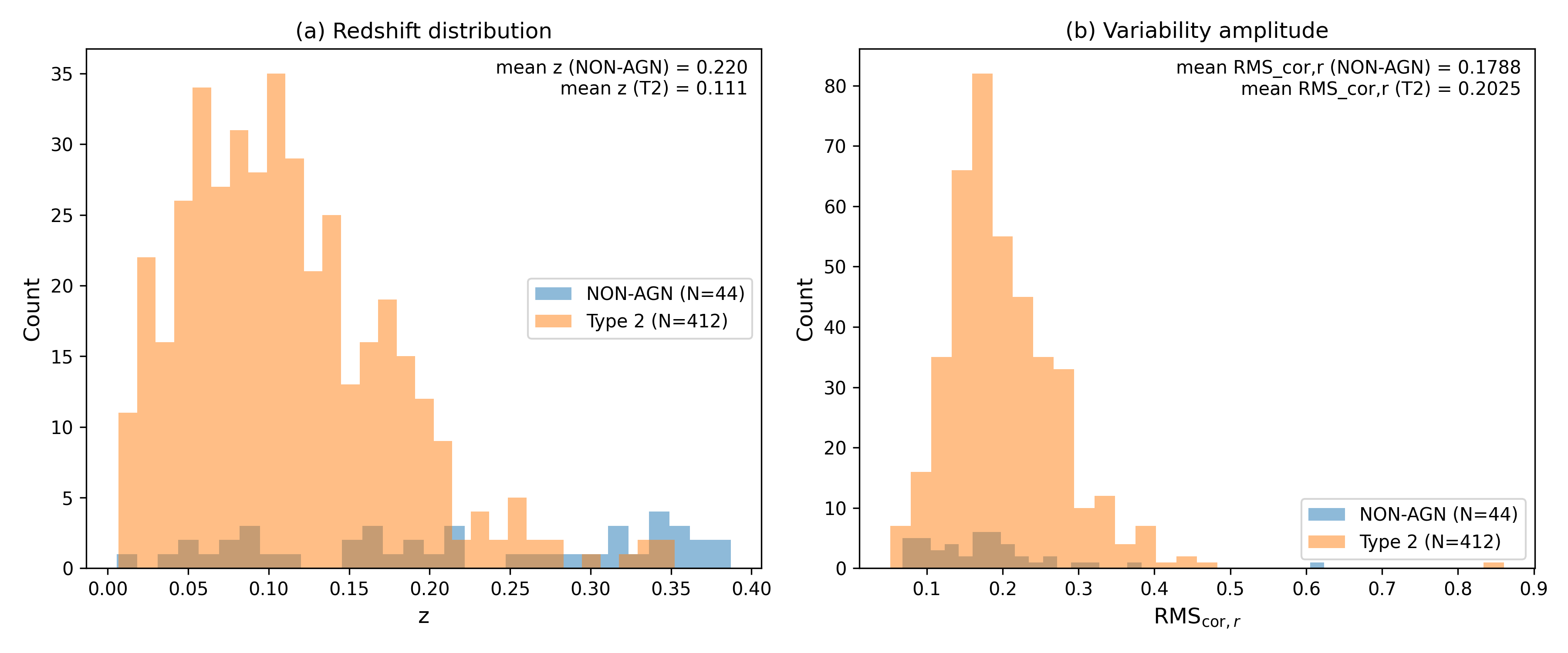}
\caption{Redshift and $\mathrm{RMS}_{\mathrm{cor},r}$ distributions for the non-AGN control sample and the full Type~2 sample.}
\label{fig:control_hist_nonagn}
\end{figure*}

For the non-AGN control sample, the noise-corrected variability amplitude, $\mathrm{RMS}_{\mathrm{cor}}$, was computed in each photometric band using the same procedure as for the Type~2 sample (Sect.~\ref{sec:variability}). Fig.~\ref{fig:control_hist_nonagn} compares the redshift and $\mathrm{RMS}_{\mathrm{cor},r}$ distributions of the non-AGN and full Type~2 samples.

The two samples differ significantly in both redshift and mean $r$-band magnitude ($\langle r\rangle$, from the \texttt{MEAN\_PSFMAG\_CLIP} column), as quantified by two-sample Kolmogorov-Smirnov tests (KS $p=1.9\times10^{-9}$ for $z$; KS $p=2.4\times10^{-3}$ for $\langle r\rangle$), indicating different apparent-flux and signal-to-noise regimes. To enable a comparison under comparable observational conditions, the Type~2 sample was therefore matched to the non-AGN control sample in redshift and $\langle r\rangle$, constructing a matched Type~2 subsample with the same number of objects as the control sample.

The matching was performed using a nearest-neighbour approach in the two-dimensional $(z,\langle r\rangle)$ parameter space. For each non-AGN object, the closest Type~2 galaxy was selected based on the Euclidean distance in standardised coordinates. Matching was performed with replacement, allowing individual Type~2 objects to be selected multiple times if they provided the closest match in parameter space.

\begin{figure*}[h!]
\centering
\includegraphics[width=180mm]{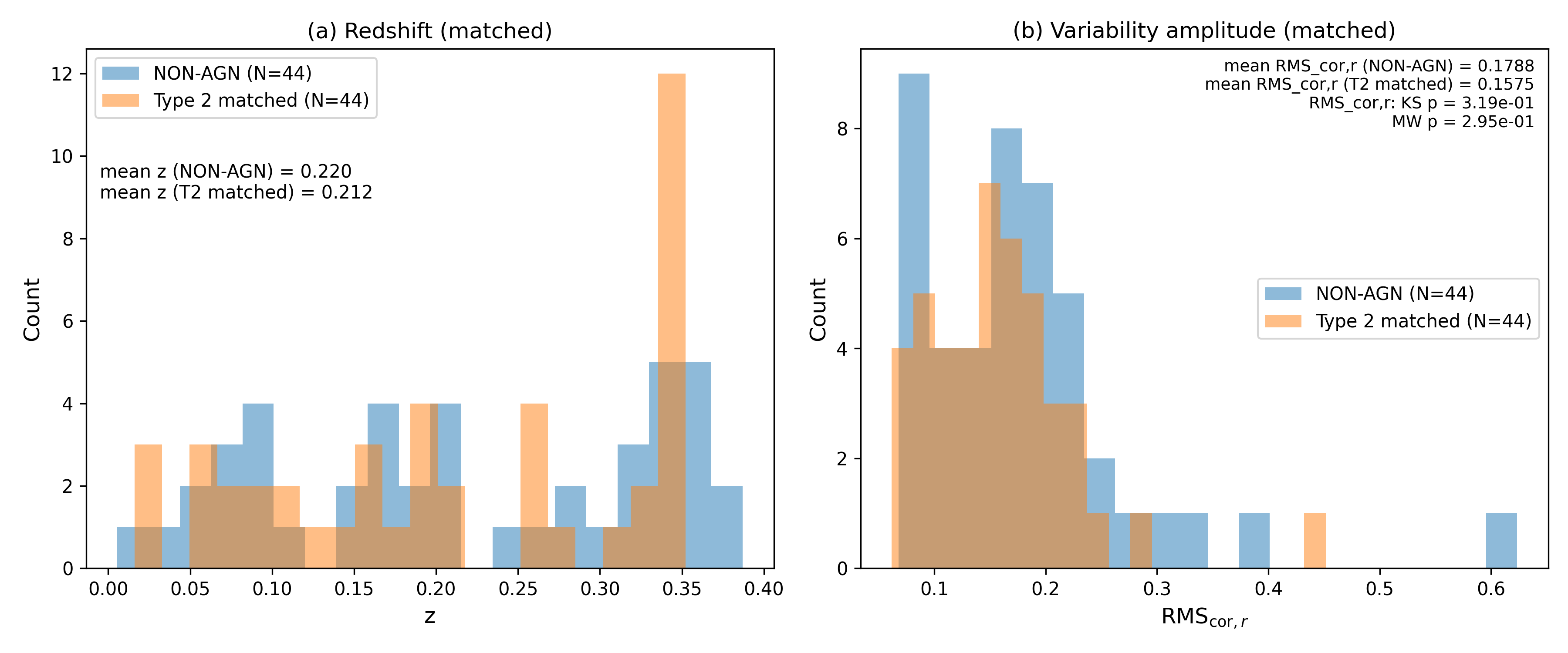}
\caption{Redshift and $\mathrm{RMS}_{\mathrm{cor},r}$ distributions for the non-AGN control sample and the matched Type~2 subsample. }
\label{fig:control_hist_nonagn_matched}
\end{figure*}

After matching, the redshift and $\langle r\rangle$ distributions of the two samples are statistically consistent (KS $p=0.81$ for both parameters). Figure~\ref{fig:control_hist_nonagn_matched} illustrates the resulting redshift and $\mathrm{RMS}_{\mathrm{cor},r}$ distributions for the matched samples. No statistically significant difference is found between their $\mathrm{RMS}_{\mathrm{cor},r}$ distributions (KS $p=0.32$; Mann--Whitney $p=0.30$), indicating comparable variability amplitudes under matched observational conditions.

\begin{figure*}[h!]
\sidecaption
\includegraphics[width=12.9cm]{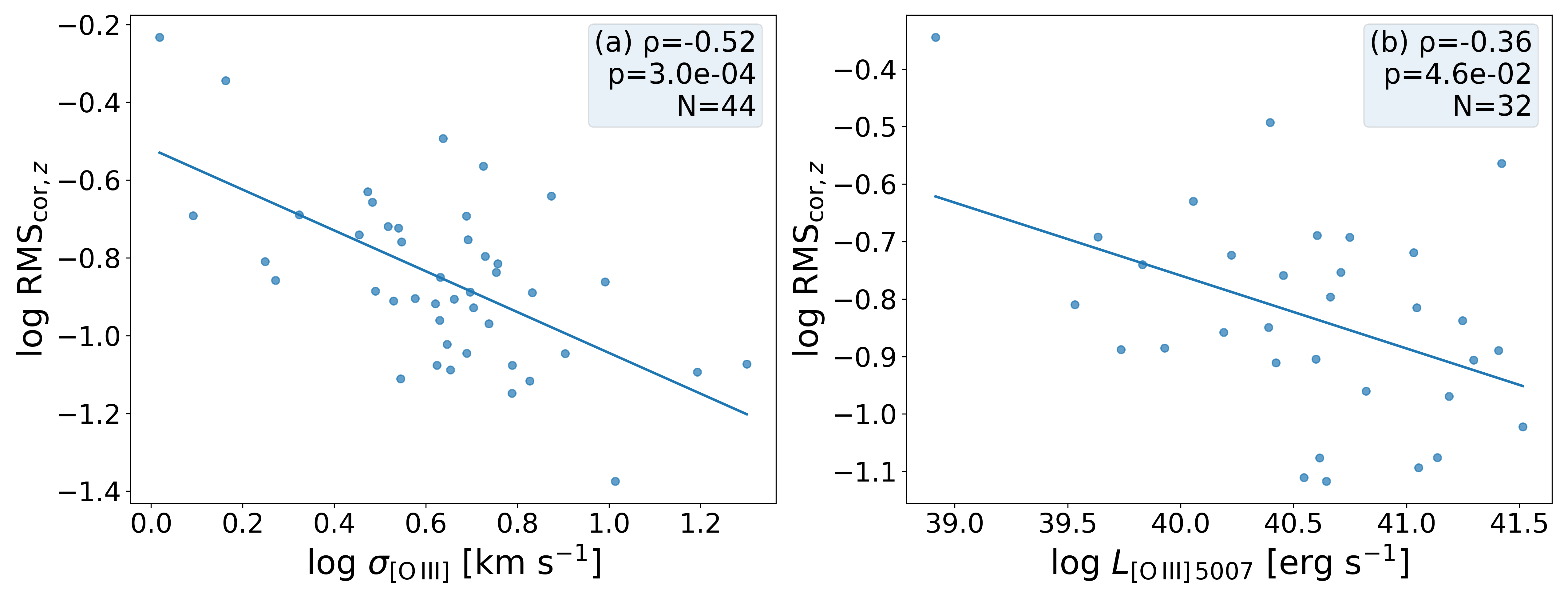}
\caption{Panel (a) shows the relation between $\mathrm{RMS}_{\mathrm{cor},z}$ and the [O\,III] velocity dispersion, while panel (b) shows the relation between $\mathrm{RMS}_{\mathrm{cor},z}$ and the [O\,III] $\lambda5007$ luminosity. These panels are intended as a control experiment to illustrate RMS variability trends that may arise from host-galaxy or observational effects.
}
\label{fig:control_nonagn_panels}
\end{figure*}

To examine whether RMS variability-spectral correlations are also present in the control sample, the same correlation analysis was performed between $\mathrm{RMS}_{\mathrm{cor}}$ and the available spectroscopic parameters for the non-AGN galaxies. All statistically significant correlations are listed in Table~\ref{tab:control}. Fig.~\ref{fig:control_nonagn_panels} shows two representative examples of RMS variability--spectral correlations in the non-AGN control sample. Given the limited size of the control sample ($N=44$), these correlations are presented for comparison purposes only.

\begin{table}
\centering
\caption{Spearman correlations between $\mathrm{RMS}_{\mathrm{cor}}$ (all bands) and spectral parameters for the non-AGN control sample.}
\label{tab:control}
\begin{tabular}{l l c c c}
\hline\hline
RMS band & Parameter & $\rho$ & $p$  & $N$ \\
\hline
RMS\_cor\_z & OIII\_4959\_FWHM & -0.56 & 1.3e-04 & 42 \\
RMS\_cor\_z & SDSS\_OIII\_SIGMA & -0.52 & 3.0e-04 & 44 \\
RMS\_cor\_i & OIII\_4959\_FWHM & -0.49 & 1.1e-03 & 42 \\
RMS\_cor\_r & OIII\_4959\_FWHM & -0.48 & 1.3e-03 & 42 \\
RMS\_cor\_g & OIII\_4959\_FWHM & -0.48 & 1.4e-03 & 42 \\
RMS\_cor\_r & SDSS\_OIII\_SIGMA & -0.45 & 2.4e-03 & 44 \\
RMS\_cor\_i & SDSS\_OIII\_SIGMA & -0.44 & 2.5e-03 & 44 \\
RMS\_cor\_g & SDSS\_OIII\_SIGMA & -0.43 & 3.5e-03 & 44 \\
RMS\_cor\_z & OIII\_5007\_FWHM & -0.39 & 1.2e-02 & 41 \\
RMS\_cor\_z & NII\_6584\_FWHM & -0.39 & 1.2e-02 & 41 \\
RMS\_cor\_z & NII\_6548\_FWHM & -0.39 & 1.2e-02 & 41 \\
RMS\_cor\_g & NII\_6548\_FWHM & -0.39 & 1.3e-02 & 41 \\
RMS\_cor\_g & NII\_6584\_FWHM & -0.39 & 1.3e-02 & 41 \\
RMS\_cor\_g & OIII\_5007\_FWHM & -0.39 & 1.3e-02 & 41 \\
RMS\_cor\_g & SDSS\_NII\_SIGMA & -0.36 & 1.8e-02 & 42 \\
RMS\_cor\_z & OIII\_5007\_LUM & -0.36 & 4.6e-02 & 32 \\
RMS\_cor\_z & SDSS\_NII\_SIGMA & -0.36 & 2.1e-02 & 42 \\
\hline
\end{tabular}
\begin{flushleft}
\textbf{Notes.} Only correlations with $|\rho|\ge 0.3$ and $p<0.05$ are listed, sorted by decreasing $|\rho|$. The column $N$ gives the number of objects used in each correlation after applying $3\sigma$ clipping.
\end{flushleft}
\end{table}

\subsection{Robustness against noise-induced selection effects} \label{sec:signed}

\begin{figure*}
\centering
\includegraphics[width=0.85\textwidth]{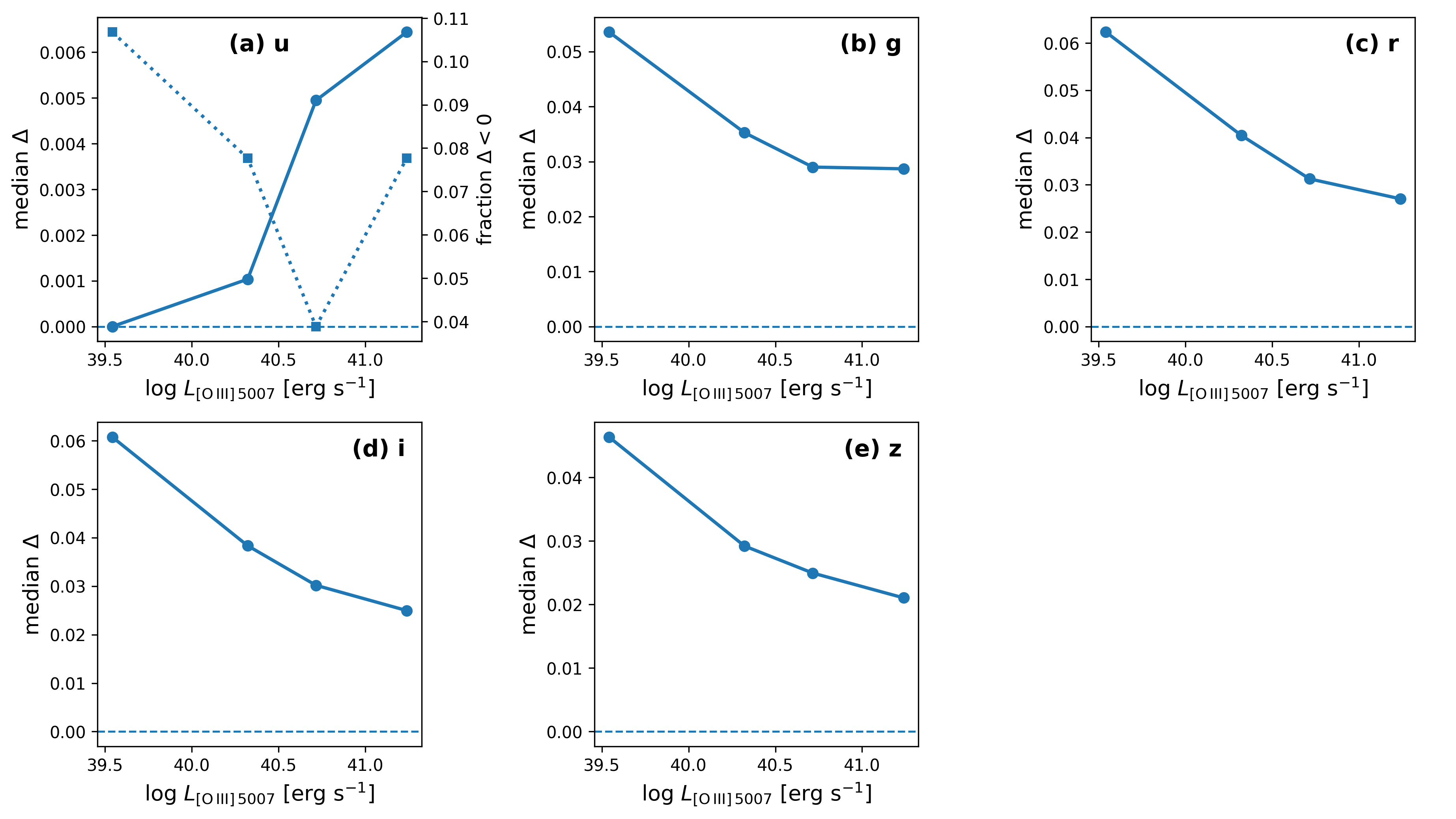}
\caption{Robustness test for noise-induced selection effects. The panels show the signed excess variance, $\Delta = \mathrm{RMS}^{2} - \langle\sigma_{\mathrm{err}}\rangle^{2}$, as a function of [O\,III] $\lambda5007$ luminosity for the $u$, $g$, $r$, $i$, and $z$ bands. Circles denote the median $\Delta$ in each luminosity bin, plotted at the median luminosity. Squares connected by dotted lines indicate the fraction of objects with $\Delta < 0$. In the $g$, $r$, $i$, and $z$ bands this fraction is zero in all bins, demonstrating that the truncation at $\mathrm{RMS}_{\mathrm{cor}}=0$ does not introduce a luminosity-dependent selection effect.}
\label{fig:delta_oiii5007}
\end{figure*}

To assess whether the truncation inherent in the definition of $\mathrm{RMS}_{\mathrm{cor}}$ (Eq.~\ref{eq:eq1}) introduces a selection bias, the analysis was repeated using the signed excess variance,
\begin{equation}
\Delta = \mathrm{RMS}^{2} - \langle\sigma_{\mathrm{err}}\rangle^{2},
\end{equation}
retaining all objects regardless of the sign of $\Delta$. The sample was divided into bins of [O\,III] $\lambda5007$ luminosity, and the median value of $\Delta$ was computed in each bin. The resulting trends (Fig.~\ref{fig:delta_oiii5007}) show the same qualitative behaviour as the $\mathrm{RMS}_{\mathrm{cor}}$ analysis, with the median $\Delta$ decreasing toward higher luminosities in the $g$, $r$, $i$, and $z$ bands, while no clear trend is observed in the $u$ band.

The fraction of objects with $\Delta < 0$ was also examined as a function of luminosity. As shown in Fig.~\ref{fig:delta_oiii5007}, this fraction is zero in all bins for the $g$, $r$, $i$, and $z$ bands, with no systematic dependence on luminosity. This demonstrates that the truncation at $\mathrm{RMS}_{\mathrm{cor}} = 0$ does not introduce a luminosity-dependent selection effect in the bands where the main correlations are observed.

A small number of objects with $\Delta < 0$ is present only in the $u$ band, reflecting the stronger impact of photometric uncertainties in this band. Consistently, no statistically significant correlation is found in the $u$ band when using the signed excess variance, indicating that the apparent RMS-based trend in this band is not robust. The interpretation of variability trends is therefore based primarily on the $g$, $r$, $i$, and $z$ bands, where the measurements are robust. These results show that the observed correlations between optical variability and emission-line properties persist when all objects are retained and are not driven by noise-induced selection effects.

\section{Discussion} \label{sec:discussion}

\subsection{Are the observed variability-spectral correlations driven by AGN activity?}

In Sect.~\ref{sec:variability}, it is shown that $\mathrm{RMS}_{\mathrm{cor}}$ in Type~2 AGNs exhibits statistically significant correlations with several narrow-line spectroscopic properties. To assess whether such trends could arise from host-galaxy systematics or observational effects, a control sample of non-AGN galaxies was constructed from the same ALPAKA--Stripe~82 parent catalogue. A matched Type~2 AGN subsample was then constructed by matching the Type~2 objects to the non-AGN control sample in redshift and $\langle r \rangle$. After matching, the two samples share statistically indistinguishable $z$ and $\langle r\rangle$ distributions, implying that observational effects related to PSF sampling, seeing, host-galaxy resolution, and photometric noise affect both samples in a comparable manner.

The non-AGN control sample nevertheless exhibits statistically significant correlations between $\mathrm{RMS}_{\mathrm{cor}}$ and several narrow-line kinematic and luminosity parameters, particularly those associated with [O\,III]. Given the absence of an active nucleus, these trends most likely reflect RMS variability components driven by host-galaxy structure, aperture effects, or measurement-related systematics. Their presence thus establishes a baseline level of RMS variability--spectral coupling that can arise even in inactive galaxies.

In contrast, the Type~2 AGN sample displays a broader and more coherent set of correlations across multiple photometric bands, involving line
luminosities, EWs, and velocity dispersions that trace both the ionisation state and the integrated kinematics of the narrow-line region. This difference suggests that the two samples do not simply differ in the overall amplitude of variability, but rather in how the RMS variability is
coupled to spectroscopic properties. Such behaviour points to an additional variability component in Type~2 AGNs that is linked to AGN-related processes.

Overall, the control-sample analysis supports the interpretation that the RMS variability--spectral correlations observed in Type~2 AGNs cannot be fully explained by host-galaxy or observational effects alone, but instead reflect an AGN-specific contribution to the observed optical variability.

\subsection{Cross-validation of line-width measurements and the origin of the RMS--$\sigma$ correlation}

\begin{figure}[h!]
\centering
\includegraphics[width=90mm]{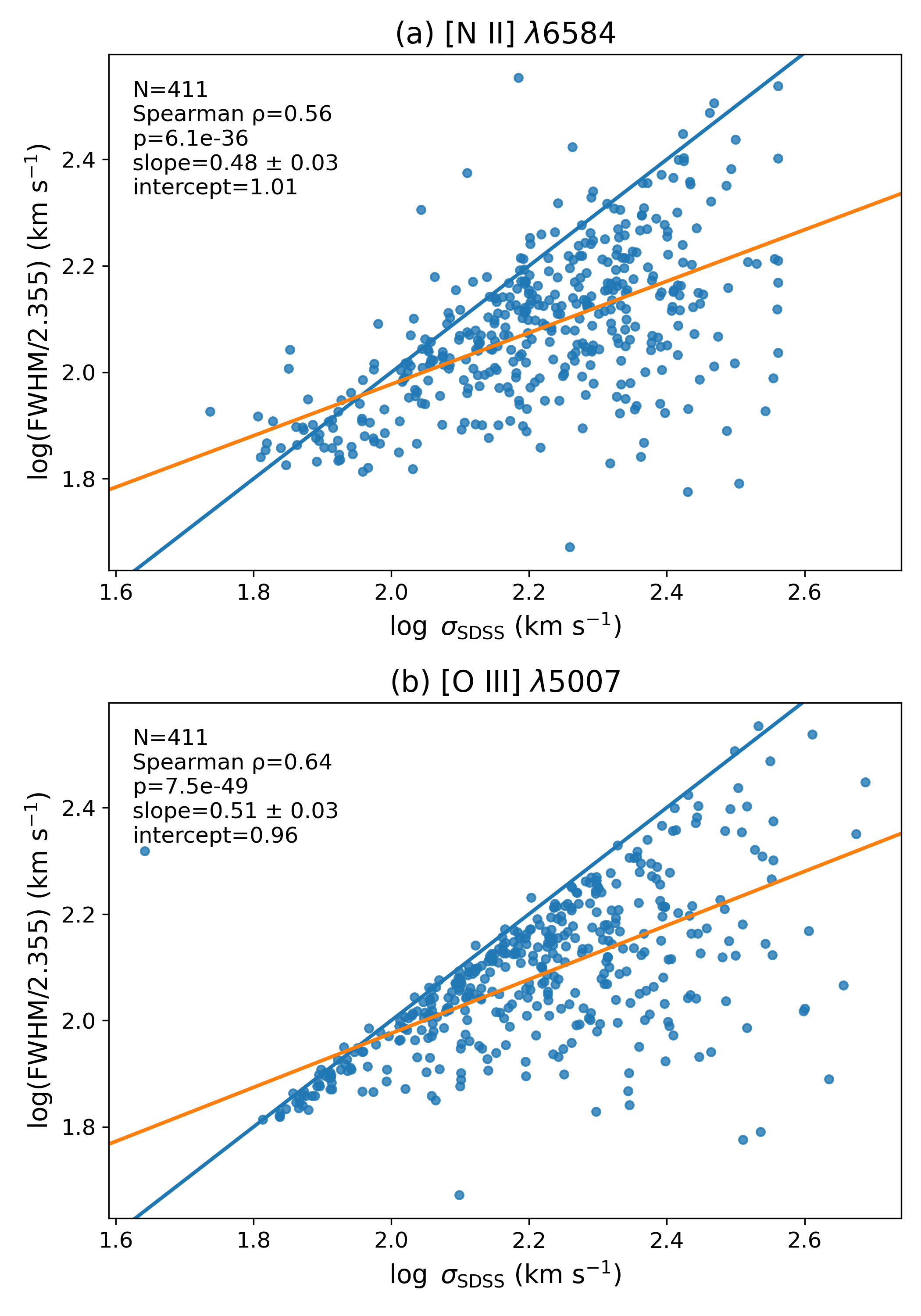}
\caption{
Cross-validation of narrow-line width measurements for Type~2 AGNs.
Panel~(a) compares $\sigma_{\rm SDSS}$ with the effective width $\mathrm{FWHM}/2.355$ for the [N\,II]~$\lambda6584$ line. Panel~(b) shows the same comparison for the [O\,III]~$\lambda5007$ line.
}
\label{fig:crossval_sigma_fwhm}
\end{figure}

The presence of a correlation between RMS$_{\mathrm{cor}}$ and line width when the latter is measured by $\sigma_{\rm SDSS}$, together with the absence of such a correlation when the same quantity is characterised by the FWHM obtained from multi-component spectral fitting, is important to investigate. Figure~\ref{fig:crossval_sigma_fwhm} presents this cross-validation for the narrow emission lines [N\,II]~$\lambda6584$ and [O\,III]~$\lambda5007$. In both cases $\sigma_{\rm SDSS}$ is compared to $\mathrm{FWHM}/2.355$, where the latter quantity represents the equivalent Gaussian dispersion. A strong and highly significant correlation is found for both lines, with Spearman coefficients $\rho\simeq0.6$.

However, the best-fitting relations deviate significantly from the one-to-one expectation. The fitted slopes are shallow ($\sim$0.5) and the intercepts are non-zero, indicating substantial intrinsic scatter and systematic differences between the two width definitions. In particular, the FWHM-based measurements are sensitive to the adopted decomposition of the line profile into individual components, whereas $\sigma_{\rm SDSS}$ captures the width of the full profile. This distinction is crucial for understanding the differing RMS trends.

As a representative example, results for the $i$ band are reported here. The observed anti-correlation between RMS$_{\mathrm{cor},i}$ and $\sigma_{\mathrm{SDSS}}$ is statistically significant for both emission lines. For [N\,II] $\lambda6584$: $\rho_{R\sigma} \simeq -0.38$, while for [O\,III] $\lambda5007$ the correlation is $\rho_{R\sigma} \simeq -0.30$.

To assess whether the RMS$_{\mathrm{cor},i}$-$\sigma_{\rm SDSS}$ relation reflects a direct physical connection or is instead induced by a third parameter, partial Spearman rank tests controlling for emission-line luminosity and EW were performed. The results are summarised in Table~\ref{tab:rms_sigma_controls}.

In this context, ``holding fixed'' $\log L$ (or $\log \mathrm{EW}$) means quantifying the rank correlation between $\log \mathrm{RMS}_{\rm cor}$ and $\log \sigma_{\rm SDSS}$ after statistically removing the monotonic dependence that each of these quantities exhibits with the chosen control parameter. For each emission line, the ordinary Spearman rank coefficients among the three logarithmic variables were computed, namely $\rho_{R\sigma}$, $\rho_{RL}$, and $\rho_{\sigma L}$ (or, alternatively, $\rho_{R\,\mathrm{EW}}$ and $\rho_{\sigma\,\mathrm{EW}}$). The standard partial-rank correlation coefficient (e.g.\ \citealt{Kendall1948}) was then evaluated as
\begin{equation}
\rho_{R\sigma\cdot L}=\frac{\rho_{R\sigma}-\rho_{RL}\,\rho_{\sigma L}}
{\sqrt{(1-\rho_{RL}^{2})(1-\rho_{\sigma L}^{2})}},
\end{equation}
with an analogous expression for $\rho_{R\sigma\cdot \mathrm{EW}}$. This statistic measures the residual correlation between RMS$_{\mathrm{cor},i}$ and line width that cannot be explained by their shared dependence on line strength. The statistical significance of the partial Spearman rank coefficients is evaluated using the standard $t$-distribution approximation for partial rank correlations, assuming $N-3$ degrees of freedom \citep{Kendall1948}.

When either the line luminosity or the EW is held fixed, the RMS$_{\mathrm{cor},i}$--$\sigma_{\rm SDSS}$ relation is no longer robust: the partial rank coefficient drops to low values and fails to meet our adopted criteria for a genuine correlation. This indicates that the apparent RMS$_{\mathrm{cor}}$--$\sigma$ anti-correlation is not intrinsic, but is largely driven by the mutual dependence of both quantities on line strength, which serves as a proxy for AGN power.

\begin{table}
\caption{Spearman and partial Spearman rank correlations for the Type~2 AGN sample.}
\label{tab:rms_sigma_controls}
\centering
\begin{tabular}{l l c c c}
\hline\hline
Emission line & Test & $\rho$ & $p$ & $N$ \\
\hline
[N\,II] $\lambda6584$
& RMS -- $\sigma_{\rm SDSS}$        & $-0.38$ & $2.37\times10^{-15}$ & 410 \\
& RMS -- $L$                        & $-0.38$ & $2.70\times10^{-15}$ & 404 \\
& $\sigma_{\rm SDSS}$ -- $L$        & $0.63$ & $1.63\times10^{-45}$ & 406 \\
& RMS -- $\mathrm{EW}$              & $-0.36$ & $2.22\times10^{-14}$ & 410 \\
& $\sigma_{\rm SDSS}$ -- $\mathrm{EW}$ & $0.67$ & $9.43\times10^{-55}$ & 412 \\
& Partial (ctrl $L$)              & $-0.19$ & $1.52\times10^{-4}$  & 404 \\
& Partial (ctrl $\mathrm{EW}$)    & $-0.19$ & $8.69\times10^{-5}$  & 410 \\
\hline
[O\,III] $\lambda5007$
& RMS -- $\sigma_{\rm SDSS}$        & $-0.30$ & $5.21\times10^{-10}$ & 409 \\
& RMS -- $L$                        & $-0.47$ & $7.77\times10^{-24}$ & 405 \\
& $\sigma_{\rm SDSS}$ -- $L$        & $0.60$ & $8.02\times10^{-41}$ & 406 \\
& RMS -- $\mathrm{EW}$              & $-0.33$ & $3.51\times10^{-12}$ & 409 \\
& $\sigma_{\rm SDSS}$ -- $\mathrm{EW}$ & $0.45$ & $9.36\times10^{-22}$ & 410 \\
& Partial (ctrl $L$)              & $0.01$ & $8.36\times10^{-1}$  & 404 \\
& Partial (ctrl $\mathrm{EW}$)    & $-0.17$ & $5.13\times10^{-4}$  & 408 \\
\hline
\end{tabular}
\begin{flushleft}
\textbf{Notes.} Spearman and partial Spearman rank correlations between $\mathrm{RMS}_{\mathrm{cor},i}$ and $\sigma_{\mathrm{SDSS}}$ are shown for the [N\,II] $\lambda6584$ and [O\,III] $\lambda5007$ emission lines. Partial correlations are computed by holding fixed the luminosity or EW of the same emission line.
\end{flushleft}
\end{table}

Overall, these tests support the interpretation that the apparent correlation between RMS$_{\mathrm{cor}}$ and $\sigma_{\rm SDSS}$ is predominantly a secondary effect. A plausible causal chain is that AGN power correlates with the integrated narrow-line width (through outflows, turbulence, and/or the host potential traced by the NLR), while RMS$_{\mathrm{cor},i}$ anti-correlates with AGN power. The combination of these two relations naturally produces an apparent RMS$_{\mathrm{cor}}$--$\sigma$ anti-correlation.

In this context, it is important to note that $\sigma_{\rm SDSS}$ reflects the width of the entire emission-line profile and is therefore sensitive to the relative contribution of low-level wings, whose prominence increases with line strength. By contrast, the FWHM values derived from multi-component fitting characterise individual kinematic components, whose widths are not required to scale monotonically with the total line luminosity. This difference naturally explains why the apparent RMS$_{\mathrm{cor}}$--$\sigma_{\rm SDSS}$ anti-correlation is observed, while no analogous trend is found when the line width is characterised by FWHM.

\subsection{Implications of the variability--spectral correlations}

The results reveal a clear anti\-correlation between RMS$_{\mathrm{cor}}$ and several narrow-line optical properties of Type~2 AGNs. Using only reliable narrow-line parameters from the ALPAKA survey (i.e.~[O\,III]~$\lambda\lambda5007,4959$ and [N\,II]~$\lambda6584$ luminosities, EWs, and kinematic widths), more variable sources tend to exhibit (i) lower forbidden-line luminosities, (ii) smaller EWs, and (iii) narrower line profiles (Table~\ref{tab:rms_trusted_corr}, Fig.~\ref{fig:panels}).

Before discussing the physical origin of these trends, their wavelength dependence is clarified. Although the correlations are strongest in the
$r$, $i$, and $z$ bands, this does not imply a larger intrinsic RMS variability amplitude at longer wavelengths: the median $\mathrm{RMS}_{\mathrm{cor}}$ is comparable across bands and does not systematically increase toward the red. In Type~2 AGNs, the weaker correlations observed in the $u$ band are likely due to stronger obscuration of the AGN continuum, such that the observed $u$-band flux may be dominated by host-galaxy emission and strongly attenuated AGN light. In addition, the robustness analysis based on the signed excess variance (Sect.~\ref{sec:signed}) shows that no statistically significant correlation is found in the $u$ band when all objects are retained. This indicates that the apparent RMS-based trend in this band is not robust and is likely related to the lower signal-to-noise ratio in the $u$ band. The main variability--spectral correlations are therefore interpreted based on the $g$, $r$, $i$, and $z$ bands, where both the $\mathrm{RMS}_{\mathrm{cor}}$ and $\Delta$ analyses yield consistent results.

Previous studies have shown that AGN optical RMS variability anti-correlates with accretion-driven parameters such as bolometric luminosity and Eddington ratio \citep[e.g.][]{SanchezSaez2018}. These relations reflect the intrinsic behaviour of the accretion disk in unobscured or partially obscured systems. In the present study of Type~2 AGNs, these anti-correlations indicate that the observed RMS and the narrow-line properties are regulated by the accretion-driven ionising output.

The relation between RMS variability and narrow-line luminosities is particularly informative in the Type~2 regime. In Type~2 AGNs, the NLR lies on sub-kiloparsec to kiloparsec scales, where the long light-travel and recombination times prevent narrow emission lines from responding to short-term continuum variability (e.g.~\citealt{Peterson1997, Netzer2013}). Consequently, the luminosities of forbidden narrow lines reflect the persistent, time-averaged ionising output of the AGN rather than the rapid month-to-year fluctuations captured by RMS variability. The fact that highly variable sources exhibit systematically lower [O\,III] and [N\,II] luminosities suggests that AGNs with strong long-term RMS variability tend to have lower time-averaged ionising output, as recorded by the narrow-line emission. This interpretation is consistent with the well-established behaviour of Type~1 AGNs, in which more luminous sources exhibit smaller amplitudes of optical variability, as demonstrated for multi-epoch light curves by \citet{Kelly2009} and further confirmed by large quasar-ensemble studies over decade-long baselines \citep{MacLeod2010}.

The correlations involving EW provide an additional constraint on the relative contribution of the AGN to the total optical light. Lower EW([O\,III]) and EW([N\,II]) values in high-RMS sources imply that, at fixed host-galaxy continuum level, the nebular emission is weaker, pointing to a lower long-term ionising photon budget or to a reduced NLR covering factor. Host-galaxy starlight can also substantially dilute both the observed continuum variability and the measured emission-line EWs. This implies that the observed trends primarily reflect the underlying AGN power \citep[e.g.][]{Kauffmann2003,Heckman2004}, with additional scatter by NLR structure and host-galaxy contamination. In Type~2 AGNs, the direct accretion-disk continuum is heavily obscured, and any observed optical continuum is generally detected only through scattered or reflected light from circumnuclear material \citep{Goodrich1995, Tran2001}. These mechanisms provide a natural pathway by which intrinsic AGN continuum variations can reach the observer despite the strong line-of-sight obscuration. Although the narrow-line flux is stable on short timescales \citep{Peterson1997,Netzer2013}, the narrow-line EW does not remain constant, because in Type~2 AGNs the observed continuum can vary through scattering, reprocessing, or variable obscuration, causing EW to respond directly to continuum-level fluctuations.

In Type~2 AGNs, forbidden-line velocity dispersions reflect both virial motions within the host-galaxy potential and additional non-virial contributions from outflows and small-scale NLR turbulence (e.g.~\citealt{NelsonWhittle1996, Mullaney2013}). More variable sources preferentially exhibit smaller $\sigma$ values for [O\,III] and [N\,II], suggesting an apparent link between optical RMS variability and the overall dynamical state of the narrow-line gas. However, the partial-rank analysis presented above shows that this RMS--$\sigma$ correlation is not primary, but instead arises as a secondary effect driven by the mutual dependence of both quantities on AGN power. The RMS variability derived from SDSS Stripe~82 traces long-term optical fluctuations over multi-year baselines (typically 6--7 years for our sample; minimum $\approx$3 years). In this regime, RMS variability is not expected to correlate directly with black hole mass \citep{Arevalo2024}, while the observed RMS--$\sigma$ anti-correlation reflects a secondary dependence. Because neither the narrow nor the broad ALPAKA FWHM components show correlations with RMS, the RMS--$\sigma$ trend is unlikely to be driven by the width of any individual Gaussian component. Each FWHM value characterises only its single fitted component, whereas $\sigma$ traces the full line-profile velocity distribution, including the relative contributions of both the core and wing emission. As a result, $\sigma$ provides a more complete measure of the combined gravitational and non-virial motions.

Some of the most optically variable sources classified as Type~2 AGNs may host very weak or highly diluted BLRs that remain undetected by automated spectral pipelines \citep{LopezNavas2023}. While such contamination cannot be fully excluded in the present sample, identifying these cases would require object-by-object spectral decomposition beyond the scope of this study.

Taken together, the anti-correlations of RMS with narrow-line luminosity, EW, and $\sigma$ suggest that optical variability in Type~2 AGNs carries information about the underlying accretion state and the structure of the circumnuclear gas, even when the BLR is completely hidden. High-RMS systems appear to correspond to less powerful AGNs, characterised by less efficient reprocessing and less kinematically disturbed NLRs. Disentangling whether the primary driver of the variability is stochastic accretion-rate fluctuations, variable obscuration, or differences in the structure of the accretion flow will require joint analyses of infrared and X-ray variability, spatially resolved spectroscopy, and detailed radiative-transfer and variability modelling. Nonetheless, these results demonstrate that continuum-driven RMS variability in obscured AGNs provides a promising, complementary probe of the long-term ionising output and the kinematic properties of the narrow-line gas.

\section{Conclusions}\label{sec:conclusions}

This work presents a study of the correlations between optical RMS variability from Stripe~82 and spectroscopic parameters from the ALPAKA survey for a sizeable, uniformly selected sample of Type~2 AGNs. These correlations enable an investigation of how the variability amplitude relates to the physical conditions of the NLR and to the obscured accretion flow. The main findings can be summarised as follows:

\begin{itemize}
  \item Optical RMS variability in Type~2 AGNs exhibits significant anti-correlations with luminosities of narrow [O\,III] $\lambda5007$, [O\,III] $\lambda4959$, [N\,II] $\lambda6584$ and [N\,II] $\lambda6548$ lines, as well as with EWs and velocity dispersion of [O\,III] $\lambda5007$ and [N\,II] $\lambda6584$. These combined trends show that AGNs with larger long-term continuum RMS variability tend to have lower narrow-line luminosities and smaller integrated line widths, both of which primarily trace the overall AGN power. Because narrow-line fluxes and kinematics probe kiloparsec-scale gas on timescales much longer than those sampled by the RMS variability, the observed RMS amplitude does not reflect changes in the NLR itself, but instead traces the underlying accretion state that governs both the NLR emission and the continuum variability.

  \item Variability amplitude decreases with narrow-line luminosity for Type~2, consistent with the well-established result that more luminous Type~1 AGNs exhibit smaller optical RMS variability.

  \item The RMS--EW anti-correlation shows that narrow-line EWs are primarily set by continuum variability: because the NLR line flux remains effectively constant on these timescales, long-term fluctuations of the obscured accretion flow change the continuum level and thereby
  alter the EW.

  \item A comparison with a matched non-AGN control sample indicates that the systematic RMS variability--spectral trends in Type~2 AGNs cannot be explained solely by host-galaxy or measurement-related effects.

\end{itemize}

Overall, these results demonstrate that optical RMS variability in Type~2 AGNs -- despite originating from a heavily obscured nucleus -- retains measurable, albeit largely secondary, connections to the properties of the NLR. The observed anti-correlations with forbidden-line luminosities and velocity dispersions primarily reflect their shared dependence on AGN power. This establishes optical RMS variability as a previously underutilized probe of the hidden AGN power, with clear links to both the long-term ionising output and the integrated kinematic state of the narrow-line gas.

\begin{acknowledgements}
This work was supported by the Ministry of Science, Technological Development and Innovations of Serbia under contract No. 451-03-33{/}2026-03{/}200002. Gratitude is expressed to the anonymous referee for constructive comments and suggestions.
\end{acknowledgements}

\appendix
\section{Impact of photometric uncertainties on variability measurements} \label{appendix}

\begin{figure}
\centering
\includegraphics[width=\columnwidth]{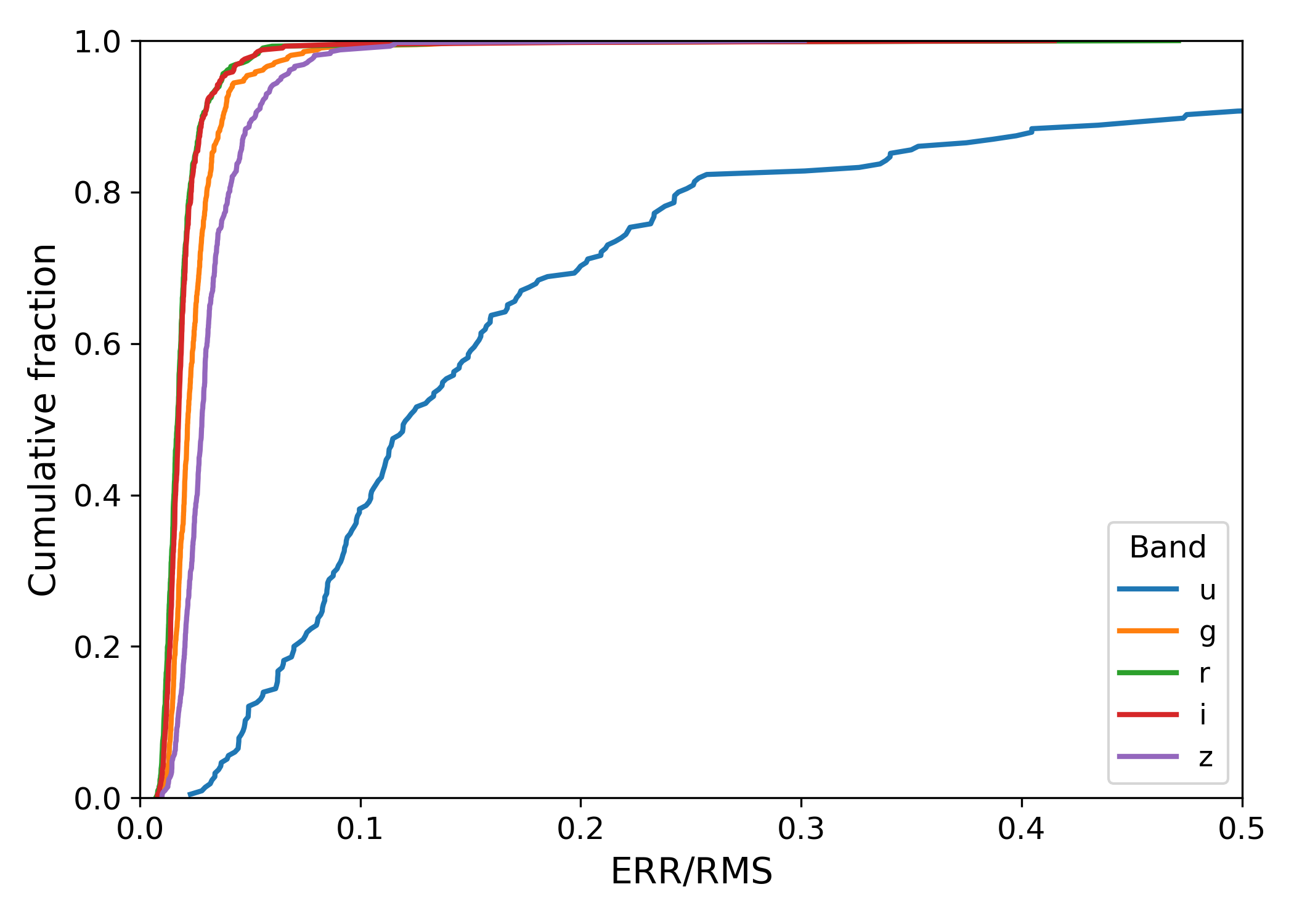}
\caption{Cumulative distributions of the $\mathrm{ERR}/\mathrm{RMS}$ ratio for the $u$, $g$, $r$, $i$ and $z$ bands. The $u$ band shows a significantly broader distribution extending to higher values, indicating a stronger impact of photometric uncertainties. In contrast, the $g$, $r$, $i$ and $z$ bands are concentrated at low $\mathrm{ERR}/\mathrm{RMS}$ values, consistent with a signal-dominated variability regime.}
\label{fig:err_rms_cdf}
\end{figure}

The impact of photometric uncertainties on the variability measurements is further assessed through the distribution of the ratio between the photometric error and the measured variability amplitude, $\mathrm{ERR}/\mathrm{RMS}$, across the $u$, $g$, $r$, $i$, and $z$ bands. Fig.~\ref{fig:err_rms_cdf} shows the cumulative distributions of $\mathrm{ERR}/\mathrm{RMS}$ for all bands. The $u$ band exhibits a significantly broader distribution extending to higher values, indicating that photometric uncertainties contribute substantially to the measured variability for a large fraction of objects. In contrast, the $g$, $r$, $i$, and $z$ bands are strongly concentrated at low $\mathrm{ERR}/\mathrm{RMS}$ values, consistent with a signal-dominated variability regime.

This behaviour is consistent with the results presented in Sect.~\ref{sec:signed}, where the signed excess variance analysis shows that variability measurements in the $u$ band are more strongly affected by photometric noise, while those in the $g$, $r$, $i$ and $z$ bands remain largely robust.

\end{document}